\documentclass[12pt]{article}
\usepackage{epsfig}
\usepackage{axodraw}
\usepackage{amssymb}
\def\beq{\begin{equation}} 
\def\eeq{\end{equation}} 
\def\bea{\begin{eqnarray}}  
\def\eea{\end{eqnarray}}  
\def\bq{\begin{quote}}  
\def\eq{\end{quote}}  
\def\bi{\begin{itemize}}  
\def\ei{\end{itemize}}  
\def\beqa{\begin{eqnarray}}  
\def\eeqa{\end{eqnarray}}  
\def\be{\begin{enumerate}}  
\def\ee{\end{enumerate}}  
\def\beq{\begin{equation}}  
\def\eeq{\end{equation}}  
\def\bi{\begin{itemize}}
\def\ei{\end{itemize}}

\def\pa{\partial}

\def\cl{{\cal L}}

\def\r2{\sqrt{2}}  

\def\bi{\begin{itemize}}  
\def\ei{\end{itemize}}

\def\ov{\overline}  
\def\nn{\nonumber \\}

\begin{document}
\pagestyle{empty}
{\normalsize\sf
\rightline {hep-ph/0109272}
\rightline{IFT-01/30}
\vskip 3mm
\rm\rightline{September 2001}
}
\begin{flushright}  

\end{flushright}
\vskip 2cm
\begin{center}
{\huge \bf Unification in models with replicated gauge groups  }
\vspace*{5mm} \vspace*{1cm} 
\end{center}
\vspace*{5mm} \noindent
\vskip 0.5cm
\centerline
{\bf Piotr H. Chankowski,  Adam Falkowski and Stefan Pokorski}
\vskip 1cm
\centerline{\em Institute of Theoretical Physics, Warsaw University}
\centerline{\em Ho\.za 69, 00-681 Warsaw, Poland}

\vskip 2cm

\centerline{\bf Abstract}
\vskip .3cm
We examine unification of gauge couplings in four dimensional renormalizable 
gauge theories inspired by the latticized (deconstructed) SM or MSSM in five dimensions. The models are based on  replicated  gauge groups, spontaneously broken to the diagonal subgroup. The analysis is performed at one-loop level, with the contribution from the heavy vector bosons included, and compared with the  analogous results in the SM or MSSM. Unification at or above the diagonal breaking scale is 
discussed. We find that in the considered class of extensions of the SM(MSSM) unification is possible for a wide range of unification scales and with the similar accuracy as in the SM(MSSM). Unification above the diagonal breaking scale is particularly attractive: it is a consequence of the SM(MSSM) unification, but with the unification scale depending on the number of replications of the gauge group.  


\newpage

\setcounter{page}{1} \pagestyle{plain}

\section{Introduction}
In recent papers it has been demonstrated \cite{AHCOGE_dec,HIPOWA} 
that there exists
an interesting class of renormalizable four dimensional gauge theories that 
predict the same  infrared
physics  as five-dimensional gauge theories. The gauge group 
of the four-dimensional theory  is a product of $N$ copies of  $SU(k)$ groups 
and is  spontaneously broken at some scale $v$ to the 
diagonal subgroup by a set
of  link-Higgs fields in the bifundamental representations 
$(\bf{k}_n, \bar{\bf{k}}_{n+1})$ of the $SU(k)_n$ group.  In addition to the massless 
gauge bosons of the unbroken $SU(k)$, the theory predicts the existence of 
massive gauge bosons  transforming in the adjoint representations of the 
diagonal $SU(k)$. In the limit of large $N$ their mass spectrum and  
interactions are identical to those of the Kaluza Klein (KK) modes of the 
compactified 5d 
$SU(k)$ gauge theory (after proper mapping of the parameters of the two 
theories; in particular the  scale $v$ of the symmetry breaking has to be 
of the same magnitude as the UV cut-off of the 5d  theory).

Thus, the four-dimensional theory at distances {\it larger} than $1/v$ 
predicts the phenomena that are usually attributed to extra dimension 
(the appearance of massive KK modes, deviations from the Coulomb law). 
The obvious advantage of the four-dimensional set-up is its renormalizability 
and, moreover, the possibility to discuss physics also at distances 
{\it shorter} than $1/v$ where the full gauge symmetry is restored. These 
properties make this class of 4d theories interesting in its own, not 
just as an UV completion of 5d non-renormalizable theories. Many  
possibilities for constructing four dimensional models inspired by extra dimensional gauge 
theories open up, and a number of examples have already been presented 
\cite{CSKRTE,CHMAWA}.

In this paper we examine the running of gauge couplings in a model based 
on  $N$-fold replication of $G_{\rm SM}\equiv SU(3)\times SU(2)\times  U(1)$ 
group of the Standard Model or its supersymmetric extension.
The theory is parametrized by N sets of gauge couplings $\alpha_1^{(n)}, \alpha_2^{(n)},\alpha_3^{(n)}$ ($n=1,\dots N$).  At some scale 
$v$
the full gauge symmetry $[G_{\rm SM}]^N$ is broken to the diagonal 
subgroup, which we identify with the SM or MSSM group with the gauge couplings  $\tilde{\alpha}_1,\tilde{\alpha}_2,\tilde{\alpha}_3$. Above the diagonal breaking scale it is natural to discuss the evolution of the couplings $\alpha_i^{(n)}$ of each single factor of  $[G_{\rm SM}]^N$, as the full gauge symmetry is restored. 
Moreover, we can  treat all the fields as massless and
the running of the gauge couplings 
proceeds as in the $\ov{MS}$ scheme. Below the 
 scale $v$ one could attempt to construct the effective theories at each stage of decoupling a single massive gauge boson but this  is problematic, since the full gauge group is broken to the diagonal subgroup in one step.  Instead, we can work  with  the full spontaneously broken  $[G_{\rm SM}]^N$ gauge theory and study the evolution of the couplings $\tilde{\alpha}_1, \tilde{\alpha}_2, \tilde{\alpha}_3$ that have the clear physical sense. At the scale $v$ the couplings $\tilde{\alpha}_i(v)$ and $\alpha_i^{(n)}(v)$ are connected to each other by the tree-level relation.  Beyond the tree-level, at scales lower than $v$ the gauge couplings of the unbroken $SU(3)\times SU(2)$ 
subgroup can be defined as the full trilinear massless gauge 
boson effective vertices \cite{POQUWE,RO}. For the unbroken $U(1)$ factor we define the gauge coupling as  the effective vertex  with the massless $U(1)$ gauge boson and two fermions.  
This allows us to include in the running
the threshold effects which are sizeable each time the RGE scale is comparable 
to the gauge bosons masses. 

In this paper we examine the evolution of the gauge couplings at one-loop level, including all contributions of the massive gauge and Higgs sectors. It is well known that in the SM and MSSM the two-loop effects are important for precision considerations and  one may expect the same to be true in the discussed class of models. Working at one-loop we cannot conclude about the absolute precision of unification but we can compare the one-loop unification  in the replicated models with the one-loop unification in the SM or MSSM. This is the goal of the present paper and we hope, that our conclusions related to such a comparison hold also at the two-loop level.  
    
One can study the gauge coupling unification at the diagonal breaking scale $v$   or above it and these two cases are qualitatively different.\footnote 
{The other logical possibility - unification below the  scale $v$ - is  very similar to the usual logarithmic running, since the heavy gauge bosons become operative not far below that scale.} 
The former 
option is similar to the power-law unification in five dimensions considered 
in \cite{DIDUGE,CHDOHI,TAVE}, but  in our case the calculation is performed 
in the renormalizable set-up and can be done rigorously.  Unification of the diagonal gauge couplings $\tilde{\alpha}_i(v)=\tilde{\alpha}(v)$ means, of course, unification of all the gauge couplings of the $[G_{SM}]^N$, $\alpha_i^{(n)}(v)=\alpha(v)$. With such a  unification scenario, one can think about the theory above the scale $v$ as e.g $[SU(5)]^N$ or string theory. Our conclusions with regard to the unification at the diagonal breaking  scale
are similar to \cite{CHDOHI}. First of all, unification is possible for a wide  range of $v < 10^{16} $GeV, but the number of replications required for unification at a  given value of $v$  is almost uniquely fixed. In the five-dimensional language it means a correlation between the cut-off scale and the compactification radius $R$. Moreover, in order to lower the unification scale sizeably we need a large number of replications.   
 In the presence of $N$  
gauge bosons the appropriate loop expansion parameter  is 
$N \tilde{\alpha} \equiv \alpha$ rather than  $\tilde{\alpha}$ itself. We find that if we 
try to lower the unification scale below $\sim 10^{10}$ GeV, then 
$\alpha$ becomes larger than 1 and the perturbative analysis cannot be 
trusted.  

Our second conclusion concerns about the comparison with one-loop unification in the SM or MSSM, which, as is well known, require threshold corrections at the unification scale of order $O(10\%)$ or  $O(1\%)$, respectively. We find that the similar conclusion holds in  the replicated SM, but in the replicated MSSM, to lower the unification scale one needs somewhat larger threshold corrections.

Unification above the diagonal breaking scale $v$ is a new interesting 
possibility in the discusssed class of  four-dimensional theories, with no counterpart in the five-dimensional theories \cite{AHCOGE_au}. In this case unification of the gauge couplings $\alpha_i^{(n)}(M_{\rm GUT})=\alpha^{(n)}(M_{\rm GUT})$ does not necessarily imply $\alpha^{(n)}(M_{\rm GUT})=\alpha(M_{\rm GUT})$.  We consider the models such that the contribution to the beta functions of any two couplings $\alpha_i^{(n)}(Q)$ and $\alpha_i^{(m)}(Q)$ can  differ by a contribution of complete representations of SU(5), e.g. matter fields. Since we are not interested in the values of the unified couplings, for our purpose the indices $(n)$ will often be in the following. The unification can take place at $M_{\rm GUT} (\gg v)$  well below $10^{16}$ GeV  and has several interesting aspects. The unification itself remains the prediction of the theory to the same extent as it is in the SM or MSSM, provided the beta functions for the couplings $\alpha_i$ above $v$ are the same as in SM(MSSM), up to full SU(5) multiplets. Furthermore, the unification takes place for a wide range of $N$, with 
$M_{\rm GUT} \approx 10^{16/N}v^{(N-1)/N}$  
and in this scenario the unification scale can be lowered sizeably even for small $N$. Therefore, contrary to the previous case of unification at the diagonal breaking scale, $v$ and $N$ remain independent parameters of the theory. Finally, at least at one-loop, the GUT scale threshold corrections needed for unification remain of the same order of magnitude as in the SM or MSSM, respectively. Thus, the models with N-fold replication of the MSSM predict unification in the same sense as does the MSSM, but with the unification scale depending on $N$.

\section{The Model}

In this section we define our set-up in which we calculate the  running of 
gauge couplings. Our guideline  
is the assumption that the success of the MSSM unification is not accidental. 
Therefore we construct our model in such a way that we replicate the gauge and Higgs structures, which are crucial for successful unification in  the MSSM. We do not replicate the matter fields which affect only the value of the gauge couplings at the unification point and not the unification itself. \footnote{Replicating matter fields would make the unified coupling $\alpha$ bigger and pertturbativity would be violated for rather small $N$.}  
 
The SM or MSSM gauge group  is replicated $N$-times. We also add $N-1$  
bifundamental link-Higgs fields $\Phi^{(n)}$ which fill out the $({\bf 5},{ \bf \bar{5}})$ 
representations of $SU(5)$. Under the $n$-th and $(n+1)$-th  
$SU(3)\times SU(2) \times U(1)$ these link fields split into  
$\Phi_{\bf 33}$ in $({\bf 3}_{1/\sqrt{15}},{\bf \bar{3}}_{-1/\sqrt{15}})$, $\Phi_{\bf 22}$ in  
$({\bf 2}_{-3/2 \sqrt{15}},{\bf \bar{2}}_{3/2 \sqrt{15}})$, $\Phi_{\bf 32}$ in 
$({\bf 3}_{1/\sqrt{15}},{\bf \bar{2}}_{3/2 \sqrt{15}})$ and $\Phi_{\bf 23}$ in 
$({\bf 2}_{-3/2 \sqrt{15}},{\bf \bar{3}}_{-1/\sqrt{15}})$.
We choose to work with the model in which 
the first and the $N$-th $SU(3)\times SU(2)\times U(1)$ group factors are not 
linked by the Higgs field.

Further, we assume that $\Phi_{\bf 33}^{(n)}$ and  $\Phi_{\bf 22}^{(n)}$ link-Higgs fields acquire 
common vacuum expectation values $\langle\Phi^{(n)}\rangle= v$. 
In the non-supersymmetric case this can be obtained  by choosing the scalar 
potential for these Higgs fields of the form:
\beq
V = \sum_n  {\rm Tr} \left[(\Phi^{(n)\;\dagger} \Phi^{(n)} - v^2)^2\right]
\eeq
In the supersymmetric case we can construct the model along the lines of 
\cite{CSGRKRTE}. The D-term potential for the link Higgs fields is:
\beq
V_D = {1\over2}g^2 \sum_{n=1}^{N}\left(
{\rm Tr}[\Phi^{(n)\;\dagger} T^a\Phi^{(n})-\Phi^{(n-1)}T^a \Phi^{(n-1)\;\dagger}]\right)^2  
\eeq
where $\Phi^{(0)}=\Phi^{(N)}=0$ and $T^a$ are the generators of the fundamental representation of the SM gauge group.  
This potential has a set of minima of the form $\Phi^{(n)}=v_n I$ where $v_n$ are its flat 
directions. Note that, with the link-Higgs fields alone, no renormalizable
superpotential can be constructed. Therefore, the flat directions of the 
potential have to be lifted by adding soft susy breaking scalar mass terms. 
Without further explanation we assume that all the VEVs are non-zero and  
equal: $v_n=v$. Expanding the link-Higgs field around the minimum,
$\Phi^{(n)}=v I + \phi^{(n)}$, we find that under the diagonal gauge group each of
the link-Higgs fields split into the following three representations: two
(real) adjoint $h_n^a= {\rm Tr}[T^a(\phi^{(n)}+(\phi^{(n)})^\dagger)]$, 
$G_n^a= i {\rm Tr}[T^a(\phi^{(n)}-(\phi^{(n)})^\dagger)]$  and the singlet 
$h_n={\rm Tr}[\phi^{(n)}]$. 
The adjoints $G_n^a$ are Goldstone bosons and become longitudinal components 
of the massive gauge bosons. The mass matrix of the physical scalars 
$h_n^a$ is identical to the one of the gauge bosons and to the mass matrices
of the Dirac fermions $\Psi_n^a$ constructed from the superpartners of the 
gauge bosons $\chi_n^a$ and the adjoint fermions coming from the superpartner 
of the link-Higgs fields $\sqrt{2} {\rm Tr}[T^a\Psi^{(n)}]$. Thus, at each gauge 
boson mass level we have a full ${\cal N}=2$ 
Yang-Mills supermultiplet. Of course, at one-loop the singlets do not affect 
the running of the gauge couplings and are irrelevant for our discussion.

The diagonal subgroup, to which the full gauge group $[G_{\rm SM}]^N$ is 
broken by the VEVs of the link-Higgs fields, is identified with the 
gauge group of the SM (MSSM). The SM (MSSM) gauge couplings $\tilde{\alpha}_i$ 
are given in terms of the couplings 
$\alpha^{(n)}_i$ of $ SU(3)_n\times SU(2)_n\times U(1)_n$ by the formula 
\begin{eqnarray}
{1\over\tilde{\alpha}_i(v)} = \sum_{n=1}^N {1\over\alpha_i^{(n)}(v)}. 
\end{eqnarray}
For the sake of simplicity we assume that the gauge couplings at the scale 
$v$ are the same for all the $N$  group factors (which is the case in the 
models of deconstructed dimensions of \cite{AHCOGE_dec,HIPOWA}) so that the 
formula for the low energy gauge couplings takes the form: 
\beq 
{1\over \tilde{\alpha}_i(v)}={N\over\alpha_i(v)}
\eeq 
The gauge bosons corresponding to the diagonal $SU(3)\times SU(2)\times U(1)$ 
subgroup remain massless (down  to the scale of electroweak symmetry breaking).
The remaining gauge boson acquire masses equal to:
\bea 
&(M_n^{(i)})^2 = 16\pi\alpha_i(v)v^2 \sin^2\left({n\pi\over2N}\right)&
\eea 
In the supersymmetric case the same masses are acquired by the corresponding Dirac gaugino and the scalar field. 

We also replicate $N$-times the {\it electroweak} Higgs doublet(s), i.e. we assume the existence
 of $N$ scalars $H_n$ transforming as $({\bf 2})_{1/2}$ of the 
$n$-th $SU(2)\times U(1)$ group. Next, we introduce trilinear couplings to 
the link-Higgs fields  of the form:
\beq 
\cl \sim  m_n (H_{n}^\dagger \Phi^{(n)}H_{n+1}) + {\rm H.c.}
\eeq
where $m_n$ are  dimensionful couplings.  
The expectation value of the link-Higgs field produces a mass matrix for 
the electroweak Higgs. We  choose the parameters 
$m_n$ so that the tower of Higgs boson
masses is the same as that of $SU(2)$ gauge bosons masses:
\beq 
\label{kkmass}
(M_n)^2 = 16\pi\alpha_2(v)v^2 \sin^2  \left({n\pi\over2N}\right)
\eeq 
In the supersymmetric case we need  $2N$ chiral supermultiplets  $H_n^{(k)}$ 
with the weak hypercharges $Y=\pm 1/2$. The mass tower similar as in the 
non-supersymmetric case can be obtained by choosing the superpotential of 
the form: 
$W\sim\sum_n(\epsilon H_n^{(2)}\Phi_{\bf 22}^{(n)} H_{n+1}^{(1)}+\mu_n H_n^{(1)}H_n^{(2)})$.
Again, by tuning the parameters of the superpotential the tower of electroweak 
Higgs mass eigenstates with masses given by (\ref{kkmass}) can be constructed.  

The matter fields, which are assumed to be those of the SM or  MSSM, are not replicated and  transform  under 
one, say the first, of the $G_{\rm SM}$ groups
\footnote{Note 
that the  running of the $\alpha_i^{(1)}$ gauge 
couplings above the scale $v$ is then different from the running of  the remaining $N-1$  couplings $\alpha^{(n)}_i$. In conseqence, the  relation 
$\alpha_i^{(1)}=\alpha_i^{(2)}=...=\alpha_i^{(N)}$ is not preserved 
above $v$. However, it still makes sense  to embed this model
at the scale $M_{\rm GUT}$
in e.g. $[SU(3)\times SU(2)\times U(1)]\times SU(5)\times\dots\times SU(5)$.}.
 The  field $\phi$ in represantation $r$ and charge $g$ of  $G_{1\; \rm SM }$, under the diagonal group transforms in the same representation with charge $\tilde{g} = {g \over \sqrt{N}}$. In addition,  $\phi$ couples to the whole tower of massive gauge bosons, and the charge is given by 
 $\tilde{g} = \sqrt{2} \cos \left({n \pi \over 2N}\right ) {g \over \sqrt{N}}$.   
\section{Unification at the diagonal breaking scale}
 
In order to define what we understand by 'successful unification' let us 
first recall the one-loop renormalization group equations in the SM and MSSM. 
At one-loop the gauge couplings $\tilde{\alpha}_i$ of the three group factors of 
$G_{\rm SM}$ run according to the equations: 
\beq 
\label{eqn:runningsm}
{1\over\tilde{\alpha}_i(Q)} = 
{1\over\tilde{\alpha}_i(M_Z)} - {b_0^{(i)}\over 2\pi} 
\ln\left({Q\over M_Z}\right)  + \delta_i  
\eeq 
Here, $1/\tilde{\alpha}_i(M_Z)=(58.98\pm0.04, ~29.57\pm0.03, ~
8.40\pm0.14)$ \cite{PDG} are the experimental values of the gauge couplings 
at the $Z^0$-pole and $b_0^{(i)}$ are the one-loop coefficients of the 
relevant beta-functions. In the limit in which all (s)particles are massless, 
they read 
$b_0=({1\over10}+{4\over3}N_g, -{43\over6}+{4\over3}N_g; -11+{4\over3}N_g)$
in the SM and $b_0=({3\over5}+2N_g; -5+2N_g; -9+2N_g)$ in the MSSM, where 
$N_g$ is the number of generations. Threshold corrections (e.g. from heavy 
GUT gauge bosons \cite{HIMUYA}) are represented by the parameters $\delta_i$.  

In the bottom-up approach one can speak about the gauge coupling unification 
if in some range of scales $Q$ the couplings defined 
by eq.~(\ref{eqn:runningsm}) with, in general, $Q$-dependent $\delta_i(Q)$ 
can take a common value $\alpha_i(Q)=\alpha_{\rm GUT}$ for reasonably small 
values of $\delta_i(Q)$ (compared to $\alpha^{-1}_{\rm GUT}$).\footnote{
Whether there exists a unified model able to provide such values of 
$\delta_i(Q)$'s is a different question.}  
The condition for the unification can be succintly written as
\beq 
\epsilon_{ijk}\left({1\over\tilde{\alpha}_i(M_Z)} + \delta_i\right)
(b_0^{(j)}- b_0^{(k)}) = 0 
\eeq
Putting in the experimental values for $\alpha_i(M_Z)$ and the 
beta-functions we get:  
\bea 
 -41.1 + 3.8 \delta_1 -11.1 \delta_2 + 7.3 \delta_3 =0 & & ({\rm SM}) 
\nn 
-0.9 +4 \delta_1 - 9.6 \delta_2 + 5.6 \delta_3 =0 & &({\rm MSSM})  
\eea

Wee see, that to achieve the gauge coupling unification at one-loop level we need the threshold 
corrections $\delta_i$ to be of order 10\% $\alpha_{\rm GUT}^{-1}$ 
in the SM, while in the MSSM we need  only $\delta_i\sim$1\%
$\alpha_{\rm GUT}^{-1}$. The latter number suggests that the gauge coupling unification in the MSSM can be very succesful. Indeed, in precision calculations 
\cite{ELKENA,CHPLPOVA} non-negligible two-loop corrections are approximately cancelled out by superpartner threshold corrections for superpartner masses O(1 TeV) \cite{CHPLPOVA}.

With the threshold corrections of the right order of magnitude, 
the  unification scale can be estimated from the equation:  
\bea 
\label{u1} 
{1\over\alpha_1(M_Z)}-{1\over\alpha_2(M_Z)}
-{1\over 2\pi}(b_0^{(1)}- b_0^{(2)})\ln\left({M_{\rm GUT}\over M_Z}\right) 
+ (\delta_1 - \delta_2) =0 
\eea 
For the sake of concreteness, in this paper we shall always  assume that $\delta_1=\delta_2=0$ and that all 
threshold corrections are accounted for by $\delta_3$ (thus, the unification 
point is assumed to be where  $\alpha_1$ and $\alpha_2$ intersect).  
Putting in the experimental numbers and the beta-function coefficients we get 
$M_{\rm GUT}\approx1\times10^{13}$ GeV
in the SM and $M_{\rm GUT}\approx 2\times10^{16}$ GeV in the MSSM.

We now turn to models with replicated gauge groups. In this section 
we investigate the possibility  that unification occurs at the  scale at which the 
$[SU(3)\times SU(2)\times U(1)]^N$ group is broken down  to its diagonal subgroup  by the Higgs boson VEV's. In practice, it is convenient to choose for this  scale 
the  mass of the heaviest gauge boson 
$M_{\rm GUT}\equiv M_{N-1}=2 g v\sin((N-1)\pi/2N)$.  The massive gauge bosons corresponding to 
broken symmetries  transform in the adjoint representation of the unbroken 
group and thus contribute to the running of the SM (MSSM) coupling constants.  
Working at one-loop level, we include them in the step-function approximation, in which  only the particles lighter than
the actual scale $Q$ contribute to the coefficients of the beta-functions. For the gauge couplings defined as full vertices, the step-function decoupling procedure is justified in Appendix A. The one-loop renormalization group equations for scales $Q$ such 
that $M_n^{(i)} < Q < M^{(i)}_{n+1}$ are 
\bea 
{1\over\tilde{\alpha}_i(Q)}={1\over\tilde{\alpha}_i(M_Z)}
- {b_0^{(i)}\over 2\pi}\ln\left({Q\over M_Z}\right)  
- {\tilde{b}^{(i)}\over 2\pi}
\ln\left({Q^n\over M_1^{(i)}\cdots M_n^{(i)}}\right)
+ \delta_i
\label{eqn:runningali}
\eea 
The coefficients of the beta-functions for the massive gauge sector 
are (see  Appendix A)
$\tilde{b}=(\frac{1}{10}, -\frac{41}{6},-\frac{21}{2})$ in the SM and  
$\tilde{b}=(\frac{3}{5}, -3,-6)$ in the MSSM. Assuming, that with the threshold corrections $\delta_i$ included, 
all $\alpha_i$ take a common value at the scale $M_{\rm GUT}$ we 
find that the KK mass levels are  the same for each gauge group: 
$M_n^{(1)}=M_n^{(2)}=M_n^{(3)}$. Therefore it is useful to define:
\beq
F_N\equiv\ln\left({M_{\rm GUT}^{N-1}\over M_1\cdots M_{N-1}}\right)
=\ln\left[{2^{N-1}\over\sqrt N}
\sin^{N-1}\left({(N-1)\pi\over2N}\right)\right]
\eeq
We see that eqs.~(\ref{eqn:runningsm}), after substituting 
$\delta_i \rightarrow \delta_i-\frac{\tilde{b}^{(i)}}{2\pi}F_N$,  
become identical to eqs.~(\ref{eqn:runningali}). In the present case 
(which is qualitatively the same as the case of extra dimensions 
\cite{DIDUGE,CHDOHI}) we have a new source of threshold corrections, with 
the function $F_N$  uniquely determined by the number of replications 
of the SM(MSSM) gauge group. The condition for unification is now:  
\beq 
\epsilon_{ijk}\left({1\over\tilde{\alpha}_i(M_Z)} + \delta_i - 
{\tilde{b}^{(i)}\over2\pi}F_N\right)\left(b_0^{(j)}- b_0^{(k)}\right) = 0 
\eeq
Putting in the experimental numbers and the coefficients of the 
beta-functions we obtain:
\bea 
\label{u3}
 -41.1 + 3.8 \delta_1 -11.1 \delta_2 + 7.3 \delta_3 + 0.01 F_N=0 & &({\rm SM}) 
\nn 
-0.9 +4 \delta_1 - 9.6 \delta_2 + 5.6 \delta_3 +0.38 F_N=0 & &({\rm MSSM})  
\eea

Let us investigate  the  impact of $F_N$ on the unification. 
With $\delta_1=\delta_2=0$, the unification scale can be 
determined from the equation: 
\bea 
\label{uscale2} 
{1\over\tilde{\alpha}_1(M_Z)}- {1\over\tilde{\alpha}_2(M_Z)} 
- {1\over2\pi}(b_0^{(1)}- b_0^{(2)})\ln\left({M_{\rm GUT}\over M_Z}\right)   
- {1\over2\pi}(\tilde{b}^{(1)}- \tilde{b}^{(2)})F_N =0 
\eea 
The results in the supersymmetric case are summarized in the Table 
\ref{tab:tab1} 
in which the required threshold correction $\delta_3$ is also given. 
It is clearly seen that in order to lower the unification scale 
significantly one needs large $N$. 
We observe also that even for the unification scale as low as $10$ TeV the 
required threshold corrections are of order 6\%. However, for unification 
scales below $\sim 10^{10}$ GeV, the effective loop expansion parameter  $\alpha \equiv N \tilde{\alpha}$ becomes larger than 1 which disfavours  
this possibility. 

\begin{table}[thb]
\caption[]{Unification at the scale $v$ in the step-function approximation 
in the supersymmetric case. \label{tab:tab1}}
\vspace{0.5cm}
\begin{center}
\begin{tabular}{|c|c|c|c|c|c|c|}
\hline
$M_{\rm GUT}$ (GeV) & $v$ (GeV) & $N$ & $F$ &
$\tilde{\alpha}_{\rm GUT}^{-1}$&$\delta_3/\tilde{\alpha}_{\rm GUT}^{-1}$&$\alpha$\\
\hline
$1.2\times10^{5} $ & $1.4\times10^{4} $ &62 & 40.2 & 47.6 & -5.4\% & 1.30\\
$1.1\times10^{6} $ & $1.3\times10^{5} $ &57 & 36.8 & 45.6 & -5.1\% & 1.25\\
$1.3\times10^{8} $ & $1.8\times10^{7} $ &46 & 29.3 & 41.3 & -4.4\% & 1.11\\
$1.1\times10^{10}$ & $1.5\times10^{9} $ &36 & 22.4 & 37.3 & -3.7\% & 0.96 \\
$1.3\times10^{12}$ & $2.1\times10^{11}$ &25 & 15.0 & 33.0 & -2.6\% & 0.76\\
$1.2\times10^{15}$ & $3.0\times10^{14}$ & 9 & 4.32 & 26.8 & -0.5\% & 0.34 \\
\hline
\end{tabular} 
\end{center}
\end{table}
\vspace{0.5cm}

In the non-supersymmetric case  the threshold corrections   
are little sensitive to  the size of $F_N$ because of the small coefficient 
in front of $F_N$ in eq.~(\ref{u3}). Thus, we can lower the unification 
scale down to $\sim10$ TeV and still maintain the threshold corrections of 
the order of  $10\%$ $\alpha_{\rm GUT}^{-1}$. { The comment on large $N$ also applies in this case.} 
 
\begin{figure}[htbp]
\begin{center}
\epsfig{file=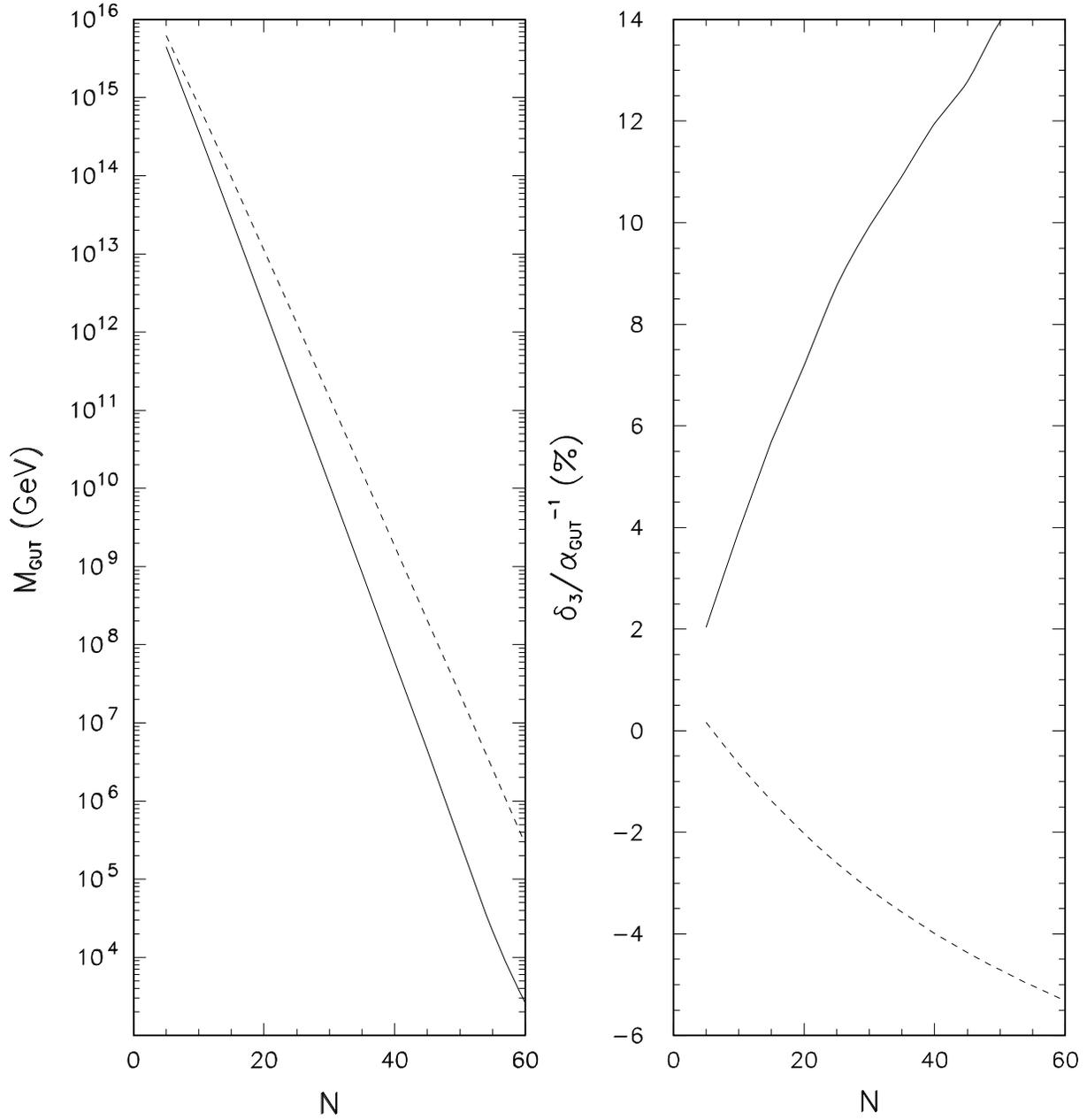,width=\linewidth}
\end{center}
\caption{\protect $M_{\rm GUT}$ and $\delta_3/\alpha_{\rm GUT}^{-1}$
(in \%) as functions of $N$ for replicated MSSM. Solid lines show the results of the 
numerical integration of the full mass-dependent beta-functions.
Dashed lines correspond to the step-function approximation.}
\label{fig:cfp1}
\end{figure}

\begin{figure}[htbp]
\begin{center}
\epsfig{file=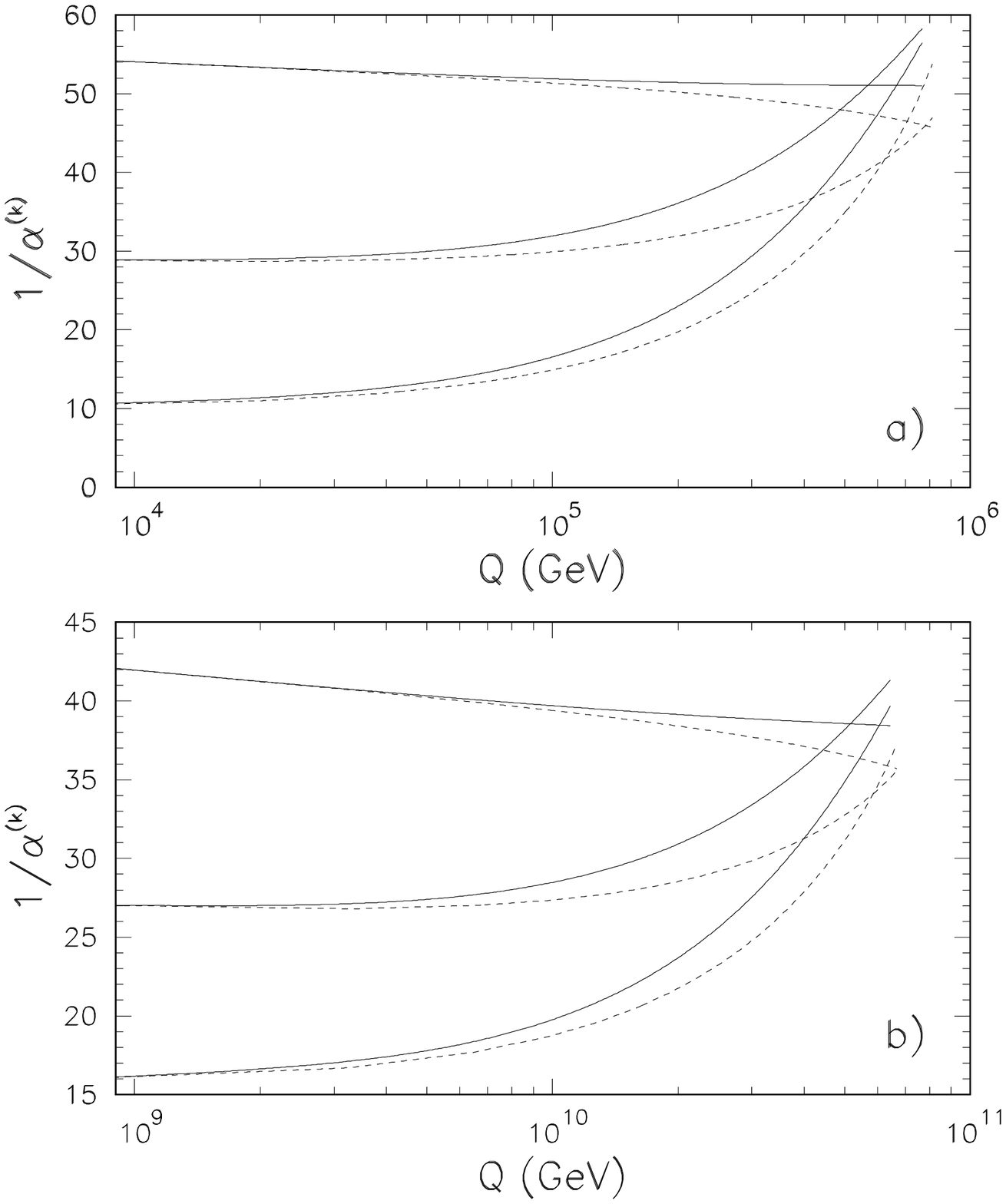,width=\linewidth}
\end{center}
\caption{\protect Evolution of (inverse) gauge couplings in replicated MSSM for  $v=10^5$ GeV, $N=60$ (panel a) and $v=10^{10}$ GeV, $N=32$ (panel b).
Solid lines show the results of the integration of the mass-dependent 
beta-functions. Dashed lines show the results obtained in the step-function
approximation.}
\label{fig:cfp2}
\end{figure}

Given the large number of the heavy vector boson thresholds it is interesting to go beyond the step-function approximation and to investigate, by means of the numerical integration, the running with the full mass-dependent beta-functions (\ref{eqn:beta}). Although these mass effects are formally of higher order, one can check they are not enhanced by the accumulation of thresholds. The results of the numerical integration are presented in  Fig.~\ref{fig:cfp1} (solid lines) and compared with those obtained in the step-function approximation. As could be expected, the difference between the two results increases with $N$ but we see, that the full mass effects are indeed of the same order of magnitude as the two-loop effects in the MSSM.  
Typical evolution of (inverse) gauge couplings in models with replicated gauge groups is shown in Fig.~\ref{fig:cfp2}a and b for $v=10^5$ GeV, $N=60$
and $v=10^{10}$ GeV, $N=32$, respectively.
 
Concluding this section,  in models with N-fold replication of the SM gauge and Higgs structures unification at the scale of symmetry breaking to the diagonal subgroup requires as large threshold corrections as the unification in the SM itself, $O(10\%)$. The threshold corrections are minimized only for strongly correlated values of $v$ and $N$. The latter remains true also for the replication of the MSSM. In this case the minimal threshold corrections required at one-loop level and in the step-function approximation remain small and for $M_{GUT}=10^{10}$ GeV
one needs $\delta_3 \sim 4\%$ for a succesful unification. The impact of the full mass dependent  beta-functions is non-negligible, but remains of the order of the known two-loop corrections in the MSSM. Thus, threshold effects of the massive gauge bosons should be considered on equal footing with two-loop corrections and other subdominant effects (e.g. e.g. corrections to the initial values of the couplings in the mass dependent renormalization scheme) in a more precise study of unification. 
The unification scale below  $10^{10}$ TeV  requires very large $N$, so that the effective loop expansion parameter $N \tilde{\alpha}$  becomes larger than one.

\section{Unification above the  scale of symmetry breaking}

If, for a fixed  $v$,  the number of replications $N$ is smaller than the value required for the unification at the scale $v$, the gauge 
couplings $\tilde{\alpha_i}$ will still differ at that  scale. 
However, above the scale $v$, 
the full symmetry group $[G_{SM}]^N$ is restored.  In our model, 
the gauge couplings $\alpha_i^{(n)}$ of all $N$ SM group factors have a common value $\alpha_i$  at the scale of the heaviest gauge boson $M_{N-1}^{(i)}$ 
and satisfy the relation:
\beq 
\label{glue}
\alpha_i(M_{N-1}^{(i)}) = N  \tilde{\alpha}_i(M_{N-1}^{(i)}) 
\eeq
in which $\tilde{\alpha}_i(M_{N-1}^{(i)})$ are determined by the running 
of gauge couplings below the diagonal breaking scale. Of course, now 
$\alpha_i(M_{N-1}^{(i)})\neq\alpha_j(M_{N-1}^{(j)})$ for
$i\neq j$. Recall that in the step-decoupling approximation:
\bea 
\label{runbelow}
{1\over\tilde{\alpha}_i(M_{N-1}^{(i)})}={1\over\tilde{\alpha}_i(M_Z)}
- {b_0^{(i)}\over 2\pi}\ln\left({M_{N-1}^{(i)}\over M_Z}\right)  
- {\tilde{b}^{(i)}\over2\pi} F_N
\eea 

Above the scale $M_{N-1}^{(i)}$  each of the gauge couplings 
$\alpha^{(i)}$ runs as in ordinary (super-)Yang-Mills theory coupled to matter.
Treating all gauge, matter and Higgs fields (including the link-Higgs fields)
as massless above that scale we have:   
\bea
\label{runover}
{1\over\alpha_i(Q)} = 
{1\over\alpha_i(M_{N-1}^{(i)})} - {\tilde{b}_0^{(i)}\over 2\pi} 
\ln\left({Q\over M_{N-1}^{(i)}}\right) + \delta_i  && (Q > M_{N-1}^{(i)})
\eea
 
The beta-function coefficients of all but the first gauge group $G_{SM}$ are 
$\tilde{b}_0 = (53/30,-11/2,-28/3)$  in the non-supersymmetric case 
 and $\tilde{b}_0 = (28/5,0,-4)$ in the supersymmetric case and include the 
 contribution from the bifundamental Higgs fields. \footnote{The coefficient of the beta-function of the first $G_{SM}$ factor gets, in addition, the 
contribution from the (MS)SM matter fields.} Combining
eqs.~(\ref{glue}), (\ref{runbelow}) and (\ref{runover}) we get: 
\bea 
\label{run}
{1\over\alpha_i(Q)}={1\over N} \left ( {1\over\alpha_i(M_Z)}
- N {\tilde{b}_0^{(i)}\over 2\pi}\ln\left({Q\over M_{N-1}^{(i)}}\right) -
 {b_0^{(i)}\over2\pi}\ln\left({Q\over M_Z} \right )  
- {\tilde{b}^{(i)}\over 2\pi} F_N  + N \delta_i \right )
\eea    

\begin{figure}[htbp]
\begin{center}
\epsfig{file=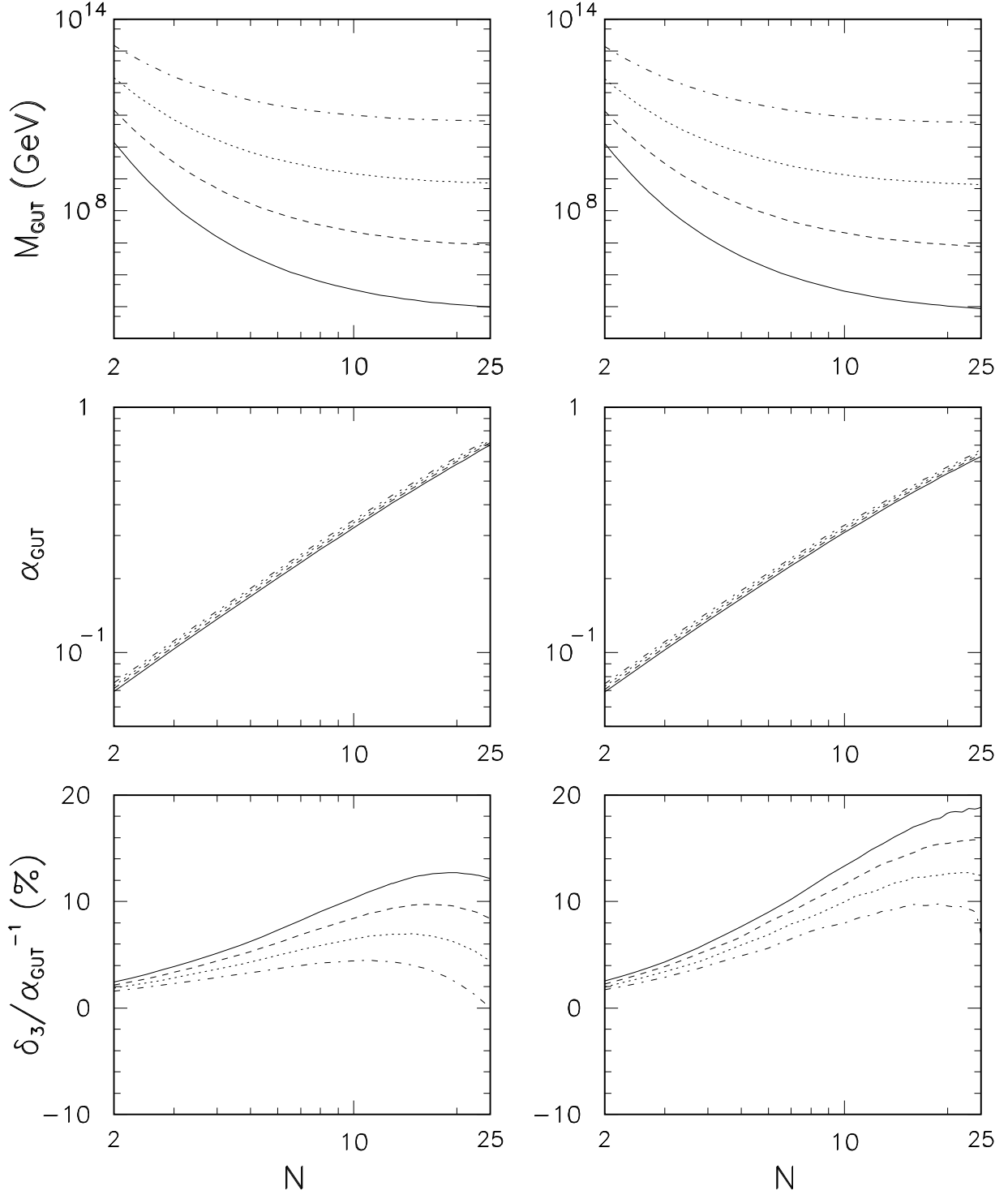,width=\linewidth}
\end{center}
\caption{\protect $M_{\rm GUT}$, $\alpha_{\rm GUT}$ and the GUT scale 
threshold corrections $\delta_3/\alpha_{\rm GUT}^{-1}$ as functions of $N$
for the unification above the diagonal breaking scale in the replicated MSSM. Left (right) panels show the results in the step-function
approximation (with the mass-dependent beta functions). Various lines correspond to $v=10^4$(solid),$10^6$(dashed),$10^8$(dotted),$10^{10}$ GeV (dash-dotted). }
\label{fig:cfp3}
\end{figure}

To obtain a transparent formula for the unification scale we need to assume 
that the heaviest gauge bosons corresponding to all the gauge factors  are 
approximately equal $M_{N-1}^{(1)}=M_{N-1}^{(2)}=M_{N-1}^{(3)}\equiv M_{N-1}$ 
(in numerical calculations, whose results are shown in the Fig.~\ref{fig:cfp3},  they usually differ by a factor of $2-5$). Then 
assuming  
$\alpha_1(M_{\rm GUT})=\alpha_2(M_{\rm GUT})=\alpha_3(M_{\rm GUT}) \equiv 
\alpha$ we get the equations: 

\bea 
\label{u2} 
{1\over\tilde{\alpha}_1(M_Z)}-{1\over\tilde{\alpha}_2(M_Z)}
-{1\over2\pi}(b_0^{(1)}- b_0^{(2)})\ln\left(
{M_{\rm U}^N\over M_{N-1}^{N-1} M_Z}\right) 
- {1\over2\pi}(\tilde{b}^{(1)}- \tilde{b}^{(2)}) F_N
+ N (\delta_1 - \delta_2) =0 
\nn
\label{u4} 
{1\over\tilde{\alpha}_2(M_Z)}-{1\over\tilde{\alpha}_3(M_Z)}
-{1\over2\pi}(b_0^{(2)}- b_0^{(3)})
\ln\left({M_{\rm U}^N\over M_{N-1}^{N-1} M_Z}\right) 
- {1\over2\pi}(\tilde{b}^{(2)}- \tilde{b}^{(3)}) F_N
+ N (\delta_2 - \delta_3) =0 
\eea  

We have used the fact that 
$\tilde{b}_0^{(i)}-\tilde{b}_0^{(j)}={b}_0^{(i)}-{b}_0^{(j)}$ 
because the bifundamental Higgs fields form complete  representations
of $SU(5)$. These  
equation have the same form as in the case of unification at the 
 scale $v$ but now \cite{AHCOGE_au}:
\bea
\label{unification_above}
M_{\rm GUT} \rightarrow {M_{\rm GUT}^N\over M_{N-1}^{N-1}}
\eea   

We also have $\delta_k \rightarrow N\delta_k$. This means that threshold 
corrections needed for succesful unification, in absolute numbers, may  be 
much smaller.  But we always express the threshold corrections in the units 
of  $\alpha^{-1}$ at the GUT scale which is also  approximetely  N-times 
smaller than in the case of unification at the  scale $v$.      
  
In Fig.~\ref{fig:cfp3} we present the results of numerical calculations in the step-decoupling approximation (left panels) and using full mass dependent beta functions (right panels). As expected from the approximate eq. (\ref{u4}), for a chosen value of $v$, the gauge couplings do unify for any number $N$ of replications. No correlation between $v$ and $N$ is needed and, in this sense, unification of the gauge couplings is the prediction of the model to the same extent as in the MSSM itself, but with $N$ dependent $M_{\rm GUT}$. A small number of replications is sufficient for reaching unification at low scale (for low $v$; of course $v$ is constrained from below $v \gtrsim 1$ TeV by experimental limits on massive vector bosons). The precision of one-loop unification is as satisfactory as in the MSSM. The effects of the full mass dependent beta-functions are very small for not too large value of $N$, say $N<10$. Moreover, for such small $N$,  $\alpha_{ \rm GUT}$ stays well within the perturbative region.        For large $N$ the couplings unify just above the diagonal breaking scale and  the unification scale and the threshold corrections are similar as in the case discussed in the previous section.  

\section{Conclusions}
We have discussed the evolution of the gauge couplings in renormalizable gauge theories based on N-fold replication of the SM(MSSM) gauge groups, spontaneously broken at some scale $v$ to the diagonal subgroup identified  with the SM(MSSM) group. A systematic one-loop calculation has been performed below and above the scale $v$. 

Unification  at the diagonal breaking scale $v$ is possible down to $v \sim 10^5$ GeV; for any value of $v$ there exists a narrow range of $N$, that gives unification at that scale. This result has the obvious counterpart in the power-law unification in five dimensions \cite{DIDUGE,CHDOHI}. 

Unification above the diagonal breaking scale is a new feature of the considered class of models \cite{AHCOGE_au}, with no counterpart in extra dimensions. It is particularly attractive as it does not require any correlation between    $v$ and $N$. It is a prediction of the model to the same extent as unification in the MSSM and is a consequence of the latter. However, in this case the unification scale is $M_{ \rm GUT} \approx v^{(N-1)/N}10^{16/N}$, and can be very low even for small $N$.

 The accuracy of one-loop unification in the considered class of models, particularly unification above the diagonal breaking scale and with not too large number of replications, is as satisfactory as in the MSSM. Of course, low unification scale raises such problems as the proton decay, but they do not rule out the idea of low scale unification \cite{DIDUGE,AHSC}.  

\vskip1.0cm
\noindent {\bf Acknowledgments}
\vskip 0.3cm
\noindent 
The work of PCH was partially supported by the EC Contract 
HPRN-CT-2000-00152 for years 2000-2004 and by the Polish State Committee 
for Scientific Research grant 5 P03B 119 20 for years 2001-2002.
The work of SP was partially supported  by the EC Contract 
HPRN-CT-2000-00148 for years 2000-2004 and by the Polish State Committee 
for Scientific Research grant 2 P03B 060 18 for years 2000-2001.
PCH would also like to thank the CERN Theory Group for hospitality during
the completion of this work.

\renewcommand{\thesection}{Appendix~\Alph{section}}
\renewcommand{\theequation}{\Alph{section}.\arabic{equation}}

\setcounter{section}{0}
\setcounter{equation}{0}

\section{}

In this appendix we specify our method of computing the RGE running of the  
gauge couplings and give the formulae for the full mass-dependent 
beta-functions which automatically take into account 
threshold effects due to the massive gauge bosons. For the 
unbroken non-abelian group factors it is convenient to define the 
 scale dependent gauge couplings as the scalar factor of the full trilinear massless vector boson vertex. 
\begin{figure}[htbp]
\begin{center}
\begin{picture}(390,100)(0,0)
\Photon(10,10)(40,50){3}{8}
\Photon(10,90)(40,50){3}{8}
\Photon(40,50)(90,50){3}{8}
\Vertex(40,50){7}
\Text(30,90)[]{$p_1$}
\Text(85,60)[]{$p_2$}
\Text(30,10)[]{$p_3$}
\Text(80,40)[]{$b,\nu$}
\Text(0,80)[]{$a,\mu$}
\Text(0,20)[]{$c,\lambda$}
\Text(250,50)[]{$V(p_1,p_2)f^{abc}g_{\mu\nu\rho}+$
other Lorentz structures}
\Text(250,30)[]
{$g_{\mu\nu\rho}=
g_{\mu\nu}(p_1 - p_2)_\rho + g_{\nu\rho}(p_2 - p_3)_\mu +g_{\rho\mu}(p_3 - p_1)_\nu$}
\end{picture}
\end{center}
\caption{1-PI three-gluon vertex.}
\label{fig:3glvert}
\end{figure}
 
At one-loop the beta-functions are given by the expression:
\beq  
\beta(q^2) = q^2{\pa\over\pa q^2}\left(2V(q^2)-3g\Pi(q^2)\right)   
\label{eqn:betadef}
\eeq
In eq.~(\ref{eqn:betadef}) $V(q^2)$ is the sum of one-loop contributions 
to the vertex defined in 
fig.~\ref{fig:3glvert} evaluated at euclidean external momenta
$p_1^2 = p_2^2 = p_3^2= -q^2<0$, and $\Pi(q^2)$ is given by the sum of the one-loop vacuum polarization diagrams contributing to the vector boson self-energy
$-i(g_{\mu\nu}p^2 - p_{\mu}p_{\nu}) \Pi(p^2)$, also evaluated at $p^2=-q^2<0$.

The vector boson self energy  gets the following contribution:
\vspace{0.5cm}
\begin{center}
\begin{tabular}{c|l}
\hline
Massless gauge boson $\matrix{\phantom{a}\cr\phantom{a}\cr\phantom{a}}$ & 
$\Pi=g^2{C_2(G)\over(4\pi)^2}\left({5\over3}B_0(m,m)-{1\over9}\right)$\\ 
\hline
Massive gauge boson $\matrix{\phantom{a}\cr\phantom{a}\cr\phantom{a}}$ & 
$\Pi=g^2{C_2(G)\over(4\pi)^2}\left({3\over2}B_0(m,m) 
-2({A(m)\over q^2} - {m^2\over q^2}B_0(m,m) + {m^2\over q^2})\right)$\\ 
\hline
Real scalar $\matrix{\phantom{a}\cr\phantom{a}\cr\phantom{a}}$ &
$\Pi= g^2{C_2(r)\over(4\pi)^2}\left(-{1\over6}B_0(m,m) 
-{2\over3}({A(m)\over q^2} 
- {m^2\over q^2}B_0(m,m) + {m^2\over q^2}) + {1\over9}\right)$\\  
\hline
Weyl fermion $\matrix{\phantom{a}\cr\phantom{a}\cr\phantom{a}}$ &
$\Pi= g^2{C_2(r)\over(4\pi)^2}\left(-{2\over3}B_0(m,m)+{4\over 3} 
({A(m)\over q^2} - {m^2\over q^2}B_0(m,m) + {m^2\over q^2})-{4\over9}\right)$\\   
\hline
\end{tabular} 
\end{center}
\vspace{0.5cm}

The vertex correction $V(q^2)$ gets the following contributions
       
\vspace{0.5cm}
\begin{center}
\begin{tabular}{c|l}
\hline
Massless gauge boson $\matrix{\phantom{a}\cr\phantom{a}\cr\phantom{a}}$ & 
$V = g^3 \frac{C_2(G)}{(4 \pi)^2} 
({2\over 3} B_0(m,m) - {1\over 36} q^2 C_0(m,m,m) -{31 \over 12})$ \\ 
\hline
Massive gauge boson $\matrix{\phantom{a}\cr\phantom{a}\cr\phantom{a}}$ & 
$V= g^3{C_2(G)\over(4\pi)^2}\left({1\over2}B_0(m,m) - 2 m^2 C_0(m,m,m)  
+ {1\over12} q^2 C_0(m,m,m) -{33\over12}\right)$\\ 
\hline
Real scalar $\matrix{\phantom{a}\cr\phantom{a}\cr\phantom{a}}$ &
$V= g^3 {C_2(r)\over(4\pi)^2}\left(-{1\over6}B_0(m,m) - {1\over 3} m^2 C_0(m,m,m)  
+ {1\over9} q^2 C_0(m,m,m) - {1\over6}\right)$\\  
\hline
Weyl fermion $\matrix{\phantom{a}\cr\phantom{a}\cr\phantom{a}}$ &
$V= g^3{C_2(r)\over(4\pi)^2}\left(-{2\over3}B_0(m,m) +{2\over3} m^2 C_0(m,m,m)  
+ {8\over9} q^2 C_0(m,m,m) - {2\over3}\right)$\\  
\hline
\end{tabular} 
\end{center}
\vspace{0.5cm}

Using the above formulae we can easily  calculate the contribution 
of the various fields to the mass dependent  beta-function. $A_\mu^{(0)}$,$A_\mu^{(n)}$,$\phi$,$\psi$ denote massless gauge boson, massive gauge boson, real scalar field and Weyl fermion, respectively.   

\vspace{0.5cm}
\begin{center}
\begin{tabular}{p{1.5cm}|l}
\hline
$A_\mu^{(0)}$ $\matrix{\phantom{a}\cr\phantom{a}\cr\phantom{a}}$ & 
$\beta = g^3 \frac{C_2(G)}{(4\pi)^2} q^2 {\pa\over\pa q^2} 
\left(-{11\over3} B_0(m,m) \right)$\\ 
\hline
$A_\mu^{(n)}$ $\matrix{\phantom{a}\cr\phantom{a}\cr\phantom{a}}$ & 
$
\beta= g^3 \frac{C_2(G)}{(4 \pi)^2}q^2 {\pa\over\pa q^2} 
\left( -{7\over 2} B_0(m,m)  
+ 6 ({A(m)\over q^2} - {m^2 \over q^2} B_0(m,m) + {m^2 \over q^2})
+(-4 m^2   + {1\over 6} q^2) C_0(m,m,m)  \right )
$
\\ 
\hline
$\phi$  $\matrix{\phantom{a}\cr\phantom{a}\cr\phantom{a}}$ &
$\beta= g^3 \frac{C_2(r)}{(4\pi)^2} q^2 {\pa\over\pa q^2}
\left( +{1\over 6} B_0(m,m) 
+ 2 ({A(m)\over q^2} - {m^2 \over q^2} B_0(m,m) + {m^2 \over q^2})
+(- {2\over 3} m^2  + {2\over 9} q^2 )C_0(m,m,m) \right)$\\  
\hline
$\psi$ $\matrix{\phantom{a}\cr\phantom{a}\cr\phantom{a}}$ & 
$\beta= g^3 \frac{C_2(r)}{(4\pi)^2} q^2 {\pa\over\pa q^2}
\left( +{2\over 3} B_0(m,m) 
- 4 ({A(m)\over q^2} - {m^2 \over q^2} B_0(m,m) + {m^2 \over q^2})
+ ({4\over 3} m^2   + {8\over 9} q^2) C_0(m,m,m)\right)$\\  
\hline
\end{tabular} 
\end{center}
\vspace{0.5cm}

$C_2(r)$ is the quadratic Casimir in  the representation $r$, by $G$ 
we denote the adjoint representation. Recall that $C_2(G)=N$ for the 
$SU(N)$ group and $C_2({\bf N})={1 \over 2}$ for the fundamental 
representation of $SU(N)$. The functions $A$, $B_0$, $C_0$ 
are defined as in \cite{AX}
\bea
&&A(m)=\int{d^4k\over(2\pi)^4}{i\over k^2 -m^2}\nonumber\\
&&B_0(p_1^2,m_1,m_2)= \int{d^4k\over(2\pi)^4}{i\over [k^2 -m^2_1][(k+p_1)^2-m_2^2]}
\nonumber\\
&&C_0(p_1,p_2,m_1,m_2,m_3)=
\int{d^4k\over(2\pi)^4}{i\over [k^2 -m^2_1][(k+p_1)^2-m_2^2]
[(k+p_1+p_2)^2-m_3^2]}\nonumber
\eea
In the formulae for the beta functions the above functions are evaluated at
 $p_1^2=p_2^2=-2p_1p_2=-
q^2$.

It is wortwhile to show that far above and far below the mass threshold our  beta functions reduce to their  ${MS}$ scheme counterparts. For $-q^2 \gg m^2$ we need the asymptotic expansions: $A(m) = m^2(-\eta - 1 + \ln(m^2))$, $B_0(m,m) = -\eta + \ln(p^2) + {\cal O}({m^2 \over q^2})$, $q^2 C_0(m,m,m) \sim  {\rm const} + {\cal O}({m^2 \over q^2})$ (where $\eta={2 \over 4-d}+\ln(4\pi)-\gamma_E$). Thus, up to  ${\cal O}({m^2 \over q^2})$ corrections, far above threshold the beta function is given by the coefficients multiplying the $B_0(m,m)$ function. The same coefficient multiplies the ${2 \over 4-d }$ singularity which yields  the  beta function in the ${MS}$ scheme. For $-q^2 \ll m^2$  we need  $B_0(m,m)= -\eta - 1 + \ln(m^2) + {\cal O}({q^2 \over m^2})$ and $C_0 \sim  {\cal O}({q^2 \over m^4})$. 
It is straightforward to see that all terms of order $ {\cal O}({m^2 \over q^2})$ and  ${\cal O}(1)$ cancel, thus far below threshold the beta function is vanishing up to  ${\cal O}({q^2 \over m^2})$ corrections.   
 
For the U(1) factor we of course do not have the trilinear gauge boson couplings. In this case it is convienient to define the scale dependent gauge coupling as the scalar factor of the full vertex with two fermions (e.g. electrons).
 
\begin{figure}[htbp]
\begin{center}
\begin{picture}(390,100)(0,0)
\Line(10,10)(40,50)
\Line(10,90)(40,50)
\Photon(40,50)(90,50){3}{8}
\Vertex(40,50){7}
\Text(30,90)[]{$p_1$}
\Text(85,60)[]{$p_2$}
\Text(30,10)[]{$p_3$}
\Text(80,40)[]{$\mu$}
\Text(0,80)[]{$$}
\Text(0,20)[]{$$}
\Text(250,50)[]{$-i V(p_1,p_2) Y \gamma_\mu +$
other Lorentz structures}
\end{picture}
\end{center}
\caption{1-PI U(1) vertex.}
\label{fig:1glvert}
\end{figure}

 The one-loop beta function is given by:
\beq  
\beta(q^2) = q^2{\pa\over\pa q^2}\left(2V(q^2)- 2 g \Sigma(q^2)- g \Pi(q^2)\right)   
\label{eqn:betau1def}
\eeq
In eq.~(\ref{eqn:betau1def}) $V(q^2)$ is the sum of one-loop contributions 
to the vertex defined in 
fig.~\ref{fig:1glvert},  $\Pi(q^2)$ is the the vector boson self-energy and $\Sigma(q^2)$ is the fermion self-energy; all diagrams being 
evaluated at euclidean external momenta
$p_1^2 = p_2^2 = p_3^2= -q^2<0$. $Y$ is the SU(5) normalized hypercharge  of the fermion.

The contributions to the $U(1)$ gauge boson self-energy are similar to the non-abelian case, with the group factors $(C_2(G),C_2(r))$ replaced by $(0,Y_r^2)$, respectively. The contributions to the fermion self-energy are:
\vspace{0.5cm}
\begin{center}
\begin{tabular}{c|l}
\hline
Massless gauge boson$\matrix{\phantom{a}\cr\phantom{a}\cr\phantom{a}}$ & 
$\Sigma=g^2{ Y^2 \over(4\pi)^2}\left(
-B_0(0,0)-1  \right )$\\ 
\hline
Massless gaugino $\matrix{\phantom{a}\cr\phantom{a}\cr\phantom{a}}$ & 
$\Sigma=g^2{ Y^2 \over(4\pi)^2}\left(
-B_0(0,0)-1  \right )$\\ 
\hline
Massive gauge boson  $\matrix{\phantom{a}\cr\phantom{a}\cr\phantom{a}}$ & 
$\Sigma=g^2{\kappa^2 Y^2 \over(4\pi)^2}\left(-B_0(m,0)-1 
-({A(m)\over q^2} - {m^2\over q^2}B_0(m,0))\right)$\\ 
\hline
Massive gaugino$\matrix{\phantom{a}\cr\phantom{a}\cr\phantom{a}}$ & 
$\Sigma=g^2{\kappa^2 Y^2 \over(4\pi)^2}\left(-B_0(m,0)-1 
+({A(m)\over q^2} - {m^2\over q^2}B_0(m,0))\right)$\\  
\hline
\end{tabular} 
\end{center}
\vspace{0.5cm}
The factors $\kappa = \sqrt{2}cos[{n \pi \over 2 N }] $ arise because the massive gauge bosons at different mass levels have different couplings to the matter fields.  $Y_r$ is the SU(5) normalized hypercharge of the (scalar/Weyl fermion) flying in the loop.  
The contributions to the  1PI vertex corrections are: 
\vspace{0.5cm}
\begin{center}
\begin{tabular}{c|l}
\hline
Massless gauge boson $\matrix{\phantom{a}\cr\phantom{a}\cr\phantom{a}}$ & 
$\Sigma=g^3{ Y^2\over(4\pi)^2}\left(
-B_0(0,0)-1  + {1 \over 3}p^2 C_0(0,0,0)\right )$\\ 
\hline
Massless gaugino$\matrix{\phantom{a}\cr\phantom{a}\cr\phantom{a}}$ &
$\Sigma=g^3{\kappa^2 Y^2 \over(4\pi)^2}
\left(- B_0(0,0) +1 
 + {2 \over 3}p^2 C_0(0,0,0))\right)$ \\  
\hline
Massive gauge boson $\matrix{\phantom{a}\cr\phantom{a}\cr\phantom{a}}$ & 
$\Sigma=g^3{\kappa^2 Y^2 \over(4\pi)^2}
\left(-{4 \over 3} B_0(m,0) +{1 \over 3} B_0(0,0) -1 
 + {1 \over 3}p^2 C_0(m,0,0))\right)$\\ 
\hline
Massive gaugino$\matrix{\phantom{a}\cr\phantom{a}\cr\phantom{a}}$ &
$\Sigma=g^3{\kappa^2 Y^2 \over(4\pi)^2}
\left(-{2\over 3} B_0(m,0) - {1 \over 3} B_0(0,0) +1 
 + ({2 \over 3}p^2 - 2 m^2) C_0(m,0,0))\right)$ \\  
\hline
\end{tabular} 
\end{center}
\vspace{0.5cm}
 Finally, we present  the contributions of various fields to the mass dependent beta-functions  of the U(1) gauge coupling.
$A_\mu^{(0)}$,$\chi^{(0)}$,$A_\mu^{(n)}$,$\chi^{(n)}$,$\phi$,$\psi$ denote massless gauge boson, massless gaugino, massive gauge boson, massive gaugino, real scalar field and Weyl fermion, respectively.    
\vspace{0.5cm}
\begin{center}
\begin{tabular}{p{1.5cm}|l}

\hline
 $A_\mu^{(0)}$ $\matrix{\phantom{a}\cr\phantom{a}\cr\phantom{a}}$ & 
$\beta = 0$\\ 
\hline
 $\chi^{(0)}$ $\matrix{\phantom{a}\cr\phantom{a}\cr\phantom{a}}$ & 
$\beta = 0$\\ 
\hline
$A_\mu^{(n)}$ $\matrix{\phantom{a}\cr\phantom{a}\cr\phantom{a}}$ & 
$\beta= g^3 \frac{Y^2 \kappa^2}{(4 \pi)^2}q^2 {\pa\over\pa q^2} 
\left( -{2\over 3} B_0 (m,0) +{2\over 3} B_0 (0,0) 
+ 2 ({A(m)\over q^2} - {m^2 \over q^2} B_0(m,0))
 + {2\over 3} q^2 C_0(m,0,0)  \right )$\\ 
\hline

$\chi^{(n)}$ $\matrix{\phantom{a}\cr\phantom{a}\cr\phantom{a}}$ & 
$\beta= g^3 \frac{Y^2 \kappa^2}{(4 \pi)^2}q^2 {\pa\over\pa q^2} 
\left( {2\over 3} B_0 (m,0) -{2\over 3} B_0 (0,0) 
- 2 ({A(m)\over q^2} - {m^2 \over q^2} B_0(m,0))
 + ({4\over 3}q^2 - 4 m^2) C_0(m,0,0)  \right )$\\ 
\hline
$\phi$  $\matrix{\phantom{a}\cr\phantom{a}\cr\phantom{a}}$ &
$\beta= g^3 \frac{ Y_r^2}{(4\pi)^2} q^2 {\pa\over\pa q^2}
\left(+{1\over 6}B_0(m,m) +{2\over3}({A(m)\over q^2} 
- {m^2\over q^2}B_0(m,m) + {m^2\over q^2})\right)$\\  
\hline
$\psi$ $\matrix{\phantom{a}\cr\phantom{a}\cr\phantom{a}}$ & 
$\beta= g^3 \frac{ Y_r^2}{(4\pi)^2} q^2 {\pa\over\pa q^2}
\left(+{2\over3}B_0(m,m)-{4\over 3} 
({A(m)\over q^2} - {m^2\over q^2}B_0(m,m) + {m^2\over q^2}) \right)$\\  
\hline
\end{tabular} 
\end{center}
\vspace{0.5cm} 

Now we are ready to write the expressions for the full beta-functions  in the supersymmetric version of our model. 
Apart from the (approximately massless) MSSM 
spectrum  with $N_g$ generations we have a tower of N-1 gauge 
multiplets of ${\cal N}=2$ supersymmetry consisting of
a vector boson, two Weyl fermions and a scalar, 
all in the adjoint represantion of the SM gauge group. The members of this multiplet have 
masses given by $M_n^{(i)}=2g(v)v\sin(n\pi/2N)$. There are also $\tilde{N_h}$ towers of 
$(N-1)$  $SU(2)$ doublet chiral supermultiplets (a scalar and a 
Weyl fermion) which for simplicity at every level are assumed to have the 
same mass as the SU(2) massive supermultiplet.

The beta-functions turn out to be:

\bea
\label{eqn:beta}
&\beta^{(k)}(q^2) = g^3{1 \over (4\pi)^2} \left( b_0^{(k)} 
+ q^2 {\pa \over \pa q^2}
\sum_{n=1}^{N-1} \tilde{b}^{(k)}(M_n^{(k)}) \right )&
\nn
&b_0^{(3)} = -9 + 2 N_g&\nn
&b_0^{(2)} = -5 + 2 N_g &\nn&
b_0^{(1)} = {3 \over 10} + 2 N_g&
\nn
&\tilde{b}^{(3)}(m)=
-6 B_0(m,m) 
- 6 m^2 C_0(m,m,m) + {13 \over 2} q^2 C_0(m,m,m) &
\nn
&\tilde{b}^{(2)}(m)=
 (-4 + {1\over2}\tilde {N}_h)B_0(m,m) 
- 4 m^2 C_0(m,m,m) 
+ ({13 \over 3}  +{2\over3}\tilde {N}_h)  
 q^2 C_0(m,m,m) &
\nn
&\tilde{b}^{(1)}(m)=
 {3\over10}\tilde {N}_h B_0(m,m)  
+  \kappa^2 (2 q^2 - 4 m^2)C_0(m,0,0) &
\eea

 In the step-decoupling approximation the contribution to the beta coefficients from the masive modes is given by the coefficient multiplying $B_0$ for $-q^2 > m^2 $ and is vanishing for $-q^2 < m^2$.

\end{document}